\documentclass[epj]{webofc}
\usepackage[utf8]{inputenc}
\usepackage[varg]{txfonts}   % Web of Conferences font
\usepackage{booktabs}
\usepackage{xcolor}
\definecolor{darkred}{rgb}{0.4,0.0,0.0}
\definecolor{darkgreen}{rgb}{0.0,0.4,0.0}
\definecolor{darkblue}{rgb}{0.0,0.0,0.4}
\usepackage[bookmarks,linktocpage,colorlinks,
    linkcolor = darkred,
    urlcolor  = darkblue,
    citecolor = darkgreen]{hyperref}
\newcommand{\p}{\partial}

\newcommand{\eqn}[1]{(\ref{#1})}
\newcommand{\wt}[1]{\widetilde{#1}}
\newcommand{\ovl}[1]{\overline{#1}}

\newcommand{\vev}[1]{\left\langle #1 \right\rangle}
\usepackage{subfigure}
\wocname{EPJ Web of Conferences}
\woctitle{Lattice2017}
\begin{document}
%%%%%%%%%%%%%%%%%%%%%%%%%%%%%%%%%%%%%%%%%%%%%%%%%%%%%%%%%%%%%%%%%%%%%%%%%%%%%
%
\selectlanguage{english}
%----------------------------------------------------------------------------
\title{%
Energy-momentum tensor correlation function in $N_f=2+1$ full QCD at finite temperature
\footnote{Talk presented at the 35th International Symposium on Lattice Field Theory (LATTICE 2017), 18-24 June 2017, Granada, Spain.}
}
%----------------------------------------------------------------------------
\author{%
\firstname{Yusuke} \lastname{Taniguchi}\inst{1}\fnsep\thanks{Speaker, \email{tanigchi@het.ph.tsukuba.ac.jp}}
\and
\firstname{Shinji} \lastname{Ejiri}\inst{2}
\and
\firstname{Kazuyuki}  \lastname{Kanaya}\inst{3}
\fnsep\thanks{present address: Tomonaga Center for the History of the Universe, University of Tsukuba, Tsukuba, Ibaraki 305-8571, Japan}
\and
\firstname{Masakiyo}  \lastname{Kitazawa}\inst{4,5}
\and
\firstname{Asobu}  \lastname{Suzuki}\inst{6}
\and
\firstname{Hiroshi}  \lastname{Suzuki}\inst{7}
\and
\firstname{Takashi}  \lastname{Umeda}\inst{8}
% etc.
% Yusuke Taniguchi, Shinji Ejiri, Kazuyuki Kanaya, Masakiyo Kitazawa, Asobu Suzuki, Hiroshi  Suzuki, Takashi Umeda
}
%----------------------------------------------------------------------------
\institute{%
Center for Computational Sciences (CCS), University of Tsukuba,
Tsukuba, Ibaraki 305-8571, Japan
\and
Department of Physics, Niigata University, Niigata 950-2181, Japan
\and
Center for Integrated Research in Fundamental Science and
Engineering (CiRfSE), University of Tsukuba, Tsukuba, Ibaraki 305-8571, Japan
\and
Department of Physics, Osaka University, Osaka 560-0043, Japan
\and
J-PARC Branch, KEK Theory Center, Institute of Particle and Nuclear Studies,
KEK, 203-1, Shirakata, Tokai, Ibaraki, 319-1106, Japan
\and
Graduate School of Pure and Applied Sciences, University of Tsukuba,
Tsukuba, Ibaraki 305-8571, Japan
\and
Department of Physics, Kyushu University, 744 Motooka, Fukuoka
819-0395, Japan
\and
Graduate School of Education, Hiroshima University,
Higashihiroshima, Hiroshima 739-8524, Japan
}
%----------------------------------------------------------------------------
\abstract{%
We measure correlation functions of the nonperturbatively renormalized energy-momentum
tensor in $N_f=2+1$ full QCD at finite temperature by applying the gradient flow method both
to the gauge and quark fields.
Our main interest is to study the conservation law of the energy-momentum tensor and
to test whether the linear response relation is properly realized for the entropy density.
By using the linear response relation we calculate the specific heat from the correlation function.
We adopt the nonperturbatively improved Wilson fermion and Iwasaki gauge action at 
a fine lattice spacing $a=0.07$ fm.
In this paper the temperature is limited to a single value $T\simeq232$ MeV.
The $u$, $d$ quark mass is rather heavy with $m_{\pi}/m_{\rho}\simeq0.63$
while the $s$ quark mass is set to approximately its physical value.
\\ \\
Preprint numbers: UTCCS-P-106, KYUSHU-HET-181, UTHEP-706, J-PARC-TH-0110}
%----------------------------------------------------------------------------
\maketitle

%----------------------------------------------------------------------------
\section{Introduction}\label{intro}

The nucleons are expected to undergo a phase transition at high temperature and density
and turn to be the quark gluon plasma (QGP).
Several heavy ion collision experiments are going on and many phenomena	have been observed
which support the transition to QGP.
One of the most fascinating observation is that of strongly coupled hydrodynamical collective motion,
which is well described with the ideal fluid model \cite{Adams:2005dq}.

At the same time there appear several challenges to calculate the viscosity of QGP
by using the lattice QCD without quarks \cite{Nakamura:2004sy,Meyer:2007ic,Meyer:2009jp}.
These pioneering works calculate the viscosity from correlation functions of the energy-momentum
tensor via the spectral function by making use of Kubo formula.
However there are two difficulties in this strategy.
One is the ill defined inverse problem to extract the continuum spectral function from
the correlation function on lattice with finite temporal extent.
The other is the renormalization of the energy-momentum tensor.
The energy-momentum tensor is the conserved current associated with the translational invariance
and is not uniquely given on lattice.
Exception is diagonal components of the tensor for the pure gluon system, for which
nonperturbative renormalization factors are given by using the thermodynamical relation.

In this paper we shall avoid the latter difficulty by applying the gradient flow 
\cite{Luscher:2010iy,Luscher:2011bx,Luscher:2013cpa}
both to the gluon and quark fields and use it as a nonperturbative renormalization scheme
\cite{Suzuki:2013gza,Makino:2014taa}.
The gradient flow renormalization scheme can be applied every component of
the energy-momentum tensor even with quarks.
The method has already been applied to the finite temperature QCD with $N_{f}=2+1$ flavors
\cite{Taniguchi:2016ofw,Taniguchi:2016tjc} and successful results are given for the equation of state
of QCD, the chiral condensate and the topological susceptibility.
The formulation is also used for a calculation of the energy-momentum tensor correlation function
in pure Yang-Mills theory at finite temperature \cite{Kitazawa:2017qab}.
We apply the strategy to $N_{f}=2+1$ flavors QCD and calculate the energy-momentum tensor
correlation function.
The purpose of this paper is to investigate (i) conservation law, (ii) restoration of rotational symmetry,
(iii) consistency between several derivation of the entropy density and (iv) specific heat.

%and plays important roles in physics;
%contains thermodynamical quantities in its diagonal components (the energy density and the pressure),
%used for a description of relativistic hydrodynamics and is a source of the gravity in the general relativity.

%----------------------------------------------------------------------------
\section{Nonperturbative renormalization using gradient flow}\label{sec-gradient-flow}

We adopt the flow equations given in Refs.~\cite{Luscher:2010iy,Luscher:2013cpa}
for the gauge and quark fields
\begin{eqnarray}
&&
   \partial_tB_\mu(t,x)=D_\nu G_{\nu\mu}(t,x),
   \quad
   B_\mu(t=0,x)=A_\mu(x),
\label{eq:(2.1)}
\\&&
\partial_t\chi_f(t,x)=\Delta\chi_f(t,x),
\quad
   \chi_f(t=0,x)=\psi_f(x),
\label{eq:(2.4)}
\\&&
\partial_t\Bar{\chi}_f(t,x)
   =\Bar{\chi}_f(t,x)\overleftarrow{\Delta},
   \quad
   \Bar{\chi}_f(t=0,x)=\Bar{\psi}_f(x),
\label{eq:(2.5)}
\end{eqnarray}
where the field strength and the covariant derivative are given in terms of the flowed gauge field
\begin{eqnarray}
&&
G_{\mu\nu}(t,x)
   =\partial_\mu B_\nu(t,x)-\partial_\nu B_\mu(t,x)
   +[B_\mu(t,x),B_\nu(t,x)],
\label{eq:(2.2)}
\\&&
   D_\nu G_{\nu\mu}(t,x)
   =\partial_\nu G_{\nu\mu}(t,x)+[B_\nu(t,x),G_{\nu\mu}(t,x)],
\label{eq:(2.3)}
\\&&
\Delta\chi_f(t,x)\equiv D_\mu D_\mu\chi_f(t,x),
\quad
   D_\mu\chi_f(t,x)\equiv\left[\partial_\mu+B_\mu(t,x)\right]\chi_f(t,x),
\label{eq:(2.6)}
\\&&
\Bar{\chi}_f(t,x)\overleftarrow{\Delta}
   \equiv\Bar{\chi}_f(t,x)\overleftarrow{D}_\mu\overleftarrow{D}_\mu,
   \quad
   \Bar{\chi}_f(t,x)\overleftarrow{D}_\mu
   \equiv\Bar{\chi}_f(t,x)\left[\overleftarrow{\partial}_\mu-B_\mu(t,x)\right].
\label{eq:(2.7)}
\end{eqnarray}
$f=u$, $d$, $s$, denotes the flavor index.

The statement of the gradient flow is that any composite operator made of flowed fields is already
renormalized and is free from the UV divergences if multiplied with the wave function
 renormalization factor for quark fields \cite{Luscher:2010iy,Luscher:2013cpa}.
The renormalization scale is given in terms of the flow time $\mu=1/\sqrt{8t}$.
The gradient flow can be used as a nonperturbative renormalization scheme when applied to the
lattice QCD Monte Carlo simulation.
An important manipulation in a nonperturbative renormalization is a conversion to commonly used
schemes like the MS or $\ovl{\rm MS}$ scheme, which are defined in perturbation theory.
This can be accomplished by two steps; (i) in the nonperturbative scheme visit a perturbative
 high energy region nonperturbatively for example by using the step scaling function,
 (ii) convert to the MS or $\ovl{\rm MS}$ scheme with a matching factor calculated perturbatively.
 
For the gradient flow scheme the first step is accomplished by following the flow towards the $t\to0$ limit,
which is realized easily in numerical simulation.
The matching factor needed for the second step is calculated in 
Refs.~\cite{Suzuki:2013gza,Makino:2014taa} for the energy-momentum tensor at one loop level.
According to the strategy the properly normalized energy-momentum tensor which satisfies
the Ward-Takahashi identity in the continuum limit is given by taking a limit
\begin{eqnarray}
T_{\mu\nu}(x)=\lim_{t\to0}T_{\mu\nu}(t,x)
\label{eqn-limit}
\end{eqnarray}
of the flowed tensor operator
\begin{eqnarray}
   T_{\mu\nu}(t,x)
   &=&\biggl\{c_1(t)\left[
   \Tilde{\mathcal{O}}_{1\mu\nu}(t,x)
   -\frac{1}{4}\Tilde{\mathcal{O}}_{2\mu\nu}(t,x)
   \right]
%\notag\\   &&\qquad{}
   +c_2(t)\left[
   \Tilde{\mathcal{O}}_{2\mu\nu}(t,x)
   -\left\langle\Tilde{\mathcal{O}}_{2\mu\nu}(t,x)\right\rangle_{\! 0}
   \right]
\notag\\   &+&
   c_3(t)\sum_{f=u,d,s}
   \left[
   \Tilde{\mathcal{O}}_{3\mu\nu}^f(t,x)
   -2\Tilde{\mathcal{O}}_{4\mu\nu}^f(t,x)
   -\left\langle
   \Tilde{\mathcal{O}}_{3\mu\nu}^f(t,x)
   -2\Tilde{\mathcal{O}}_{4\mu\nu}^f(t,x)
   \right\rangle_{\! 0}
   \right]
\notag\\   &+&
  c_4(t)\sum_{f=u,d,s}
   \left[
   \Tilde{\mathcal{O}}_{4\mu\nu}^f(t,x)
   -\left\langle\Tilde{\mathcal{O}}_{4\mu\nu}^f(t,x)\right\rangle_{\! 0}
   \right]
%\notag\\   &&\qquad{}
   +\sum_{f=u,d,s}c_5^f(t)\left[
   \Tilde{\mathcal{O}}_{5\mu\nu}^f(t,x)
   -\left\langle\Tilde{\mathcal{O}}_{5\mu\nu}^f(t,x)\right\rangle_{\! 0}
   \right]\biggr\},
\label{eq:(2.8)}
%\nn\\
\end{eqnarray}
where $\langle\cdots\rangle_{0}$ stands for the vacuum expectation value
(VEV), i.e., the expectation value at zero temperature.
The flowed tensor operator is given by a linear combination of five operators $\tilde{O}_{i\mu\nu}(t,x)$
%\begin{align}
%   \Tilde{\mathcal{O}}_{1\mu\nu}(t,x)&\equiv
%   G_{\mu\rho}^a(t,x)\,G_{\nu\rho}^a(t,x),
%\label{eq:(2.9)}
%\\
%   \Tilde{\mathcal{O}}_{2\mu\nu}(t,x)&\equiv
%   \delta_{\mu\nu}\,G_{\rho\sigma}^a(t,x)\,G_{\rho\sigma}^a(t,x),
%\label{eq:(2.10)}
%\\
%   \Tilde{\mathcal{O}}_{3\mu\nu}^f(t,x)&\equiv
%   \varphi_f(t)\,\Bar{\chi}_f(t,x)
%   \left(\gamma_\mu\overleftrightarrow{D}_\nu
%   +\gamma_\nu\overleftrightarrow{D}_\mu\right)
%   \chi_f(t,x),
%\label{eq:(2.11)}
%\\
%   \Tilde{\mathcal{O}}_{4\mu\nu}^f(t,x)&\equiv
%   \varphi_f(t)\,\delta_{\mu\nu}\,
%   \Bar{\chi}_f(t,x)
%   \overleftrightarrow{\Slash{D}}
%   \chi_f(t,x),
%\label{eq:(2.12)}
%\\
%   \Tilde{\mathcal{O}}_{5\mu\nu}^f(t,x)&\equiv
%   \varphi_f(t)\,\delta_{\mu\nu}\,
%   \Bar{\chi}_f(t,x)\,
%   \chi_f(t,x)
%\label{eq:(2.13)}
%\end{align}
with matching coefficients $c_{i}(t)$ given in Refs.~\cite{Suzuki:2013gza,Makino:2014taa} using MS
scheme
\footnote{In Ref.~\cite{Taniguchi:2016ofw} the coefficients calculated in $\ovl{\rm MS}$ scheme is used.}.
%Note 
%\begin{equation}
%   \overleftrightarrow{D}_\mu\equiv D_\mu-\overleftarrow{D}_\mu,
%\label{eq:(2.14)}
%\end{equation}
%and $\varphi_f(t)$ is the wave function renormalization factor (squared) of the quark field
%given by~\cite{Makino:2014taa},
%\begin{equation}
%   \varphi_f(t)\equiv
%  - \frac{6}
%   {(4\pi)^2\,t^2
%   \left\langle\Bar{\chi}_f(t,x)\overleftrightarrow{\Slash{D}}\chi_f(t,x)
%   \right\rangle_{\! 0}}.
%\label{eq:(2.15)}
%\end{equation}

We notice that five operators are expected to be evaluated on lattice and the continuum limit is taken
before the $t\to0$ limit in \eqn{eqn-limit}.
The $t\to0$ limit is necessary in order to resolve a mixing with irrelevant dimension six operators
which is proportional to the flow time $t$.

%----------------------------------------------------------------------------
\section{Observables}\label{sec-observables}

In this paper we shall calculate two point correlation functions of the fluctuation
of the energy-momentum tensor
\begin{eqnarray}
 C_{\mu\nu;\rho\sigma}(\tau)&=&
\frac{1}{V_3T^5}\int_{V_3}d^3xd^3y
\vev{\delta T_{\mu\nu}(\tau,\vec{x})\delta T_{\rho\sigma}(0,\vec{y})}_T,
\end{eqnarray}
where
\begin{eqnarray}
\delta T_{\mu\nu}(\tau,\vec{x})&=&
T_{\mu\nu}(\tau,\vec{x})-\vev{T_{\mu\nu}(\tau,\vec{x})}_T
\end{eqnarray}
and $\tau$ is the Euclidean time.
$\langle\cdots\rangle_{T}$ is the one point function at the same temperature.

The renormalized energy-momentum tensor consists of gluon contribution given by the first line of
\eqn{eq:(2.8)} and that from quarks given by the second and third lines in \eqn{eq:(2.8)}.
In two point correlation functions, they lead to contributions categorized into two types;
(i) $\vev{({\rm gluon})\cdot({\rm gluon})}$, $\vev{({\rm gluon})\cdot({\rm quark})}$ and
disconnected $\vev{({\rm quark})\cdot({\rm quark})}$ two point functions,
(ii) connected $\vev{({\rm quark})\cdot({\rm quark})}$ two point functions.
The first terms are calculated by using the same technique used in Ref.~\cite{Taniguchi:2016ofw},
where the noise estimator is adopted for the quark one point functions by inserting the noise vector
at flow time $t$.
In this paper we put the noise vector at flow time $t=0$ in order to save the numerical cost.

For the second contribution from connected quark diagrams we need the flowed quark propagator
\cite{Luscher:2013cpa}.
\begin{eqnarray}
\vev{\chi_r(t,x)\Bar{\chi}_r(t,y)}_F
=\wt{S}_r(t,x;t,y)-c_{\rm fl}\wt{C}(t,x;t,y),
\label{propagator}
\end{eqnarray}
where $c_{\rm fl}$ is the $O(a)$ improvement factor and we adopt $c_{\rm fl}=1/2$ at tree level.
$\wt{S}$ and $\wt{C}$ are given by
\begin{eqnarray}
&&
\wt{S}_{f}(t,x;t,y)=\sum_{vw}K(t,x;0,v)S_f(v,w)K(t,y;0,w)^\dagger,
\\&&
\wt{C}(t,x;t,y)=\sum_{vw}K(t,x;0,v)\delta_{vw}K(t,y;0,w)^\dagger,
\end{eqnarray}
where $v$ and $w$ stands for a apace-time point at $t=0$.
$S_f(v,w)$ is the bare quark propagator at $t=0$.
$K$ and $K^{\dagger}$ is the flow kernel which satisfies the ordinary flow equation
\begin{eqnarray}
&&
\left(\p_{t}-\Delta_{x}\right)K(t,x;s,y)=0,
\quad
K(t,x;t,y)=\delta_{x,y},
\label{normal-flow}
\end{eqnarray}
and the adjoint flow equation \cite{Luscher:2013cpa}
\begin{eqnarray}
&&
\left(\p_{s}+\Delta_{y}\right)K(t,x;s,y)^{\dagger}=0
\quad
K(t,x;t,y)^{\dagger}=\delta_{x,y}.
\label{adjoint}
\end{eqnarray}
In an evaluation of the propagator \eqn{propagator} we set a point source on $y$ at flow time $t$
and solve the adjoint flow equation \eqn{adjoint} to $t=0$, where the inverse of the Dirac operator
is calculated.
Starting from a source $S_f(v,w)K(t,y;0,w)^\dagger$ for $v$ at $t=0$ we solve the flow equation
\eqn{normal-flow} and get the propagator.
Numerical implementation of both flow equation is already given in
 Refs.~\cite{Taniguchi:2016ofw,Taniguchi:2016tjc}.

%----------------------------------------------------------------------------
\section{Numerical results}\label{sec-results}

Measurements of the energy-momentum tensor are performed on $N_f=2+1$ gauge
configurations generated for Ref.~\cite{Umeda:2012er}.
As given in \eqn{eq:(2.8)} we need to subtract the zero-temperature values of the operators.
The zero temperature gauge configurations are also prepared which were generated
for Ref.~\cite{Ishikawa:2007nn}.

The nonperturbatively $O(a)$-improved Wilson quark action and
the renormalization-group improved Iwasaki gauge action are adopted.
The bare coupling constant is set to $\beta=2.05$, which corresponds
to~$a=0.0701(29)\,\mathrm{fm}$ ($1/a\simeq2.79\,\mathrm{GeV}$).
The hopping parameters are set to $\kappa_u=\kappa_d\equiv\kappa_{ud}=0.1356$
and $\kappa_s=0.1351$, which correspond to heavy $u$ and $d$ quarks,
$m_\pi/m_\rho\simeq0.63$, and almost physical $s$ quark, $m_{\eta_{ss}}/m_\phi\simeq0.74$.
See Ref.~\cite{Taniguchi:2016ofw} for a detailed explanation of numerical parameters.
The temperature is limited to a single value $T\simeq232$ MeV.

%Parameters for the numerical simulation are given in Table \ref{tab-1}.
%\begin{table}[thb]
%  \small
%  \centering
%  \caption{Parameters for the numerical simulation: Temperature in MeV,
%$T/T_{\mathrm{pc}}$ assuming $T_{\mathrm{pc}}=190$ MeV, the temporal lattice size $N_t$
%and the number of configurations used in gauge and fermion measurements. 
%}
%  \label{tab-1}% Give a unique label
%  \begin{tabular}{ccccc}\toprule
% $T$[MeV] & $T/T_{\mathrm{pc}}$ & $N_t$ & gauge confs. & fermion confs. \\
%\midrule
% $0$ & $0$ & $56$ & $650$ & $65$ \\
% $232$ & $1.22$ & $12$ & $1290$ & $129$ \\
%\bottomrule
%  \end{tabular}
%\end{table}

In this section we perform three tests for the conservation law, the rotational symmetry
 and the linear response relations.
We calculate the specific heat by using the linear response relation.

%----------------------------------------------------------------------------
\subsection{Conservation law}
\label{sec:conservation}

\begin{figure}[thb] % no figure before 1st section
  \centering
  \includegraphics[width=4.5cm]{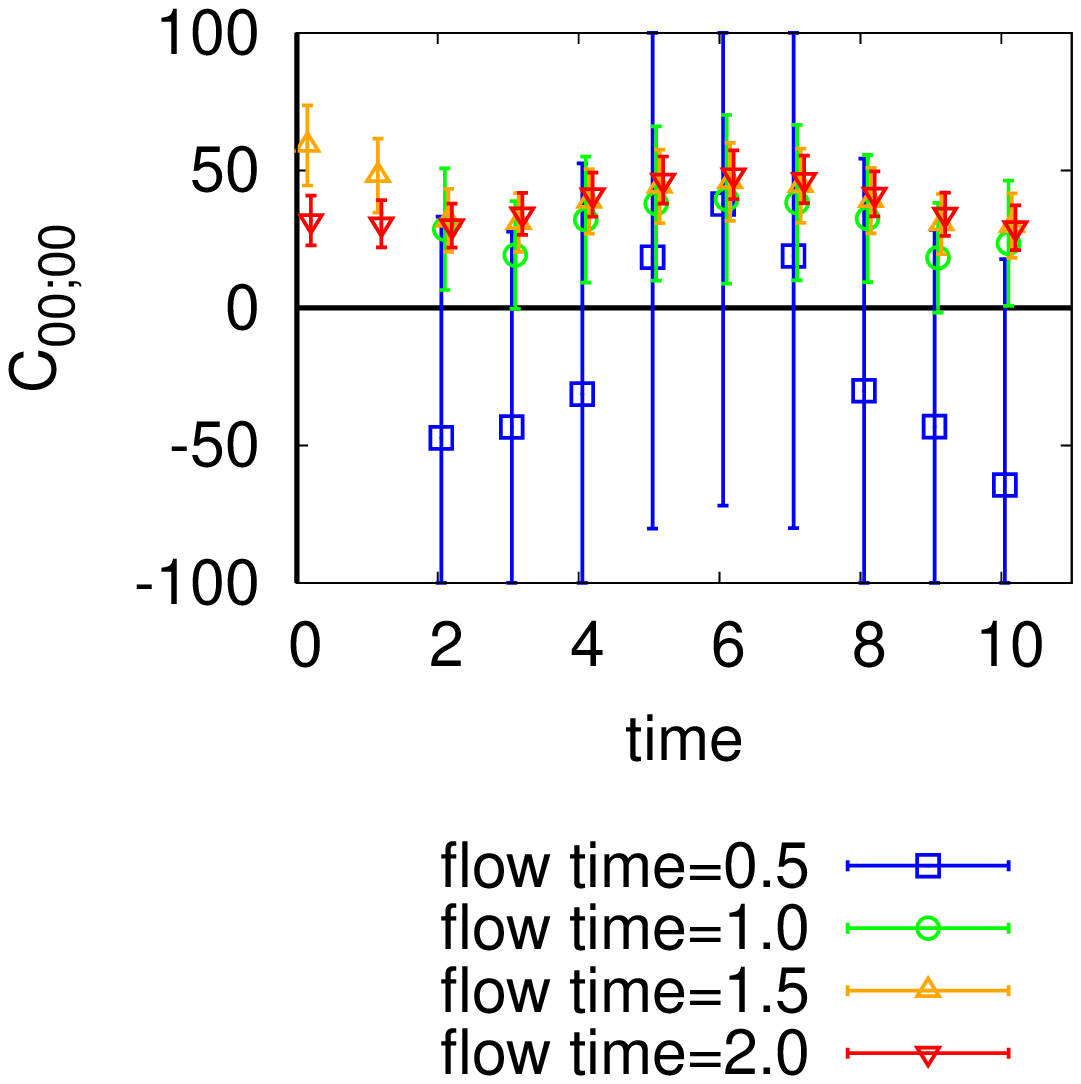}
  \includegraphics[width=4.5cm]{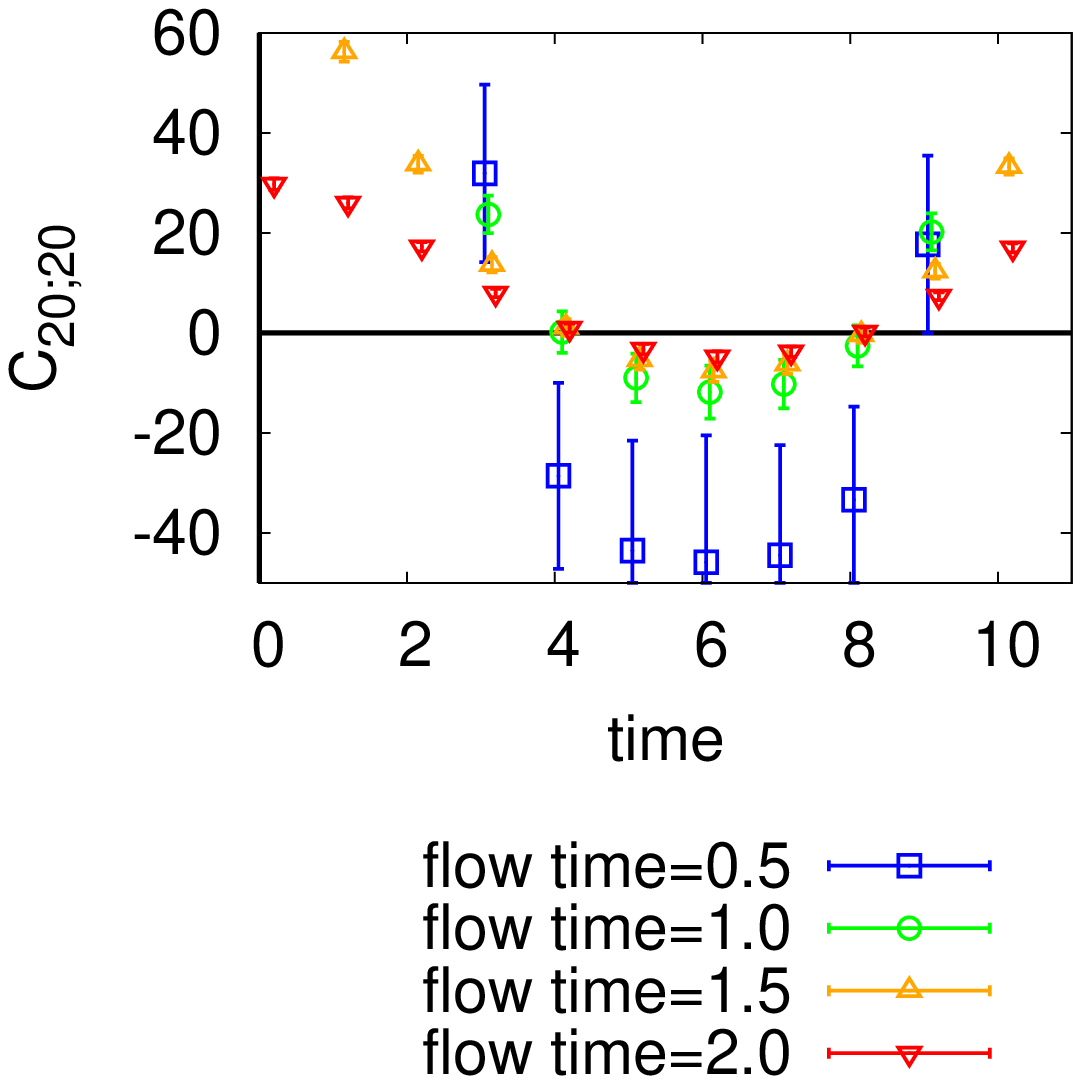}
  \includegraphics[width=4.5cm]{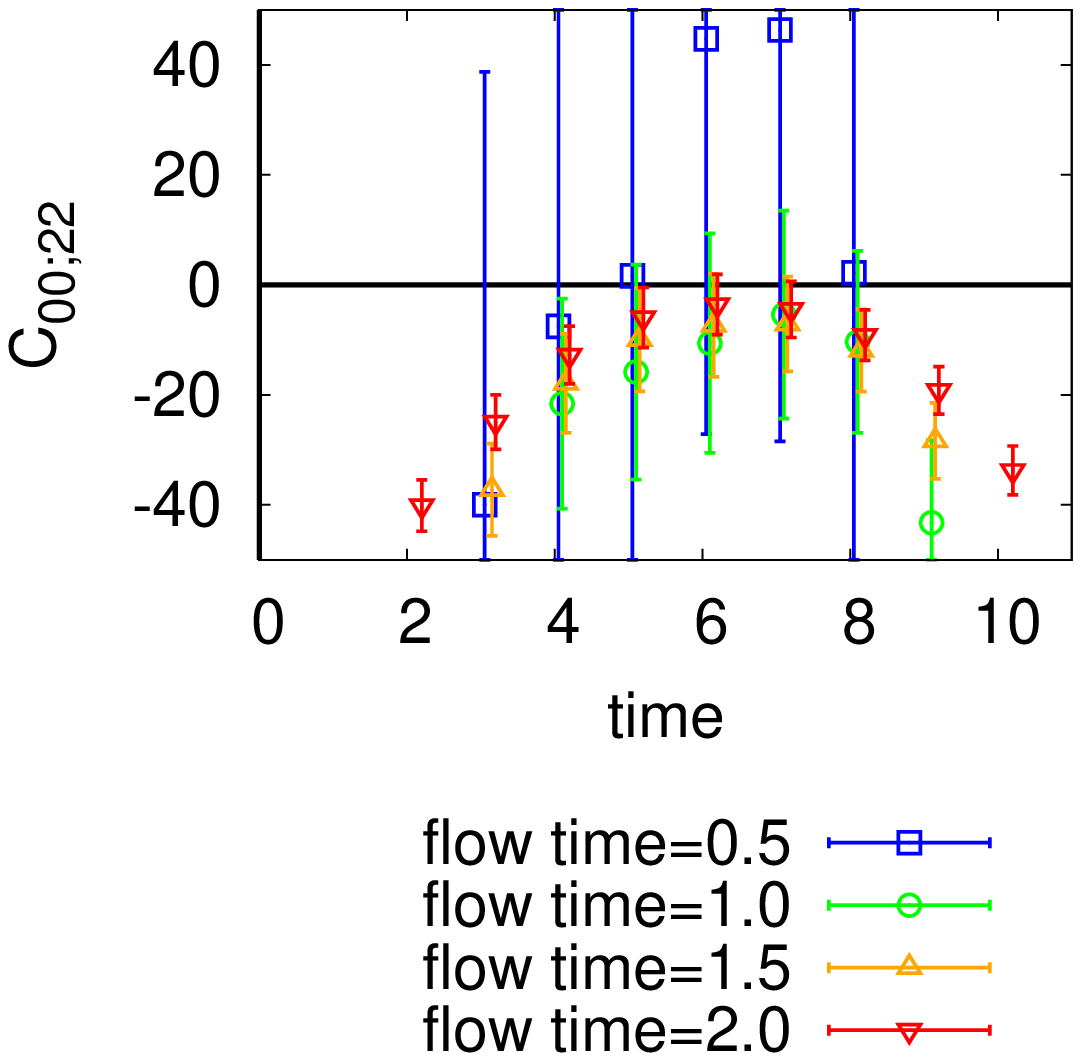}
  \caption{Correlation functions $C_{00;00}$ (left), $C_{20;20}$ (middle) and $C_{00;22}$ (right)
   as a function of the Euclidean time at flow time $t/a^{2}=0.5$ (blue square), $1.0$ (green circle), 
  $1.5$ (orange up triangle), $2.0$ (red down triangle).}
  \label{fig:conservation}% Give a unique label
\end{figure}

Spatial integral of temporal component of the energy-momentum tensor is a conserved charge
in the continuum limit and satisfies the conservation law
\begin{eqnarray}
\frac{d}{dx_{0}}C_{0\nu;\rho\sigma}(x_{0})=0.
\end{eqnarray}
Three corresponding correlation functions $C_{00;00}$, $C_{20;20}$ and $C_{00;22}$ are
plotted as a function of the Euclidean time in Fig.~\ref{fig:conservation} at flow time
$t/a^{2}=0.5, 1.0, 1.5, 2.0$.
As the gradient flow goes ahead statistical quality of the correlation function is improved.
In the middle of the Euclidean time $5\le\tau/a\le7$ the function is consistent with flat within one sigma.
%with high statistical precision for $t/a^{2}\ge1.5$,
The flat plateau is expected to spread as we approach the continuum limit and adopt larger $N_{t}$
in temporal direction as is shown in Ref.~\cite{Kitazawa:2017qab}.
If this is the case we regard the plateau as a realization of the conservation law on lattice.
On the other hand the correlation function is far from being flat near the boundary where we
 set the point source.
This is considered to be due to the $a^{2}/\tau^{2}$ lattice artifact.

%----------------------------------------------------------------------------
\subsection{Rotational symmetry}

\begin{figure}[thb] % no figure before 1st section
  \centering
  \includegraphics[width=5.4cm]{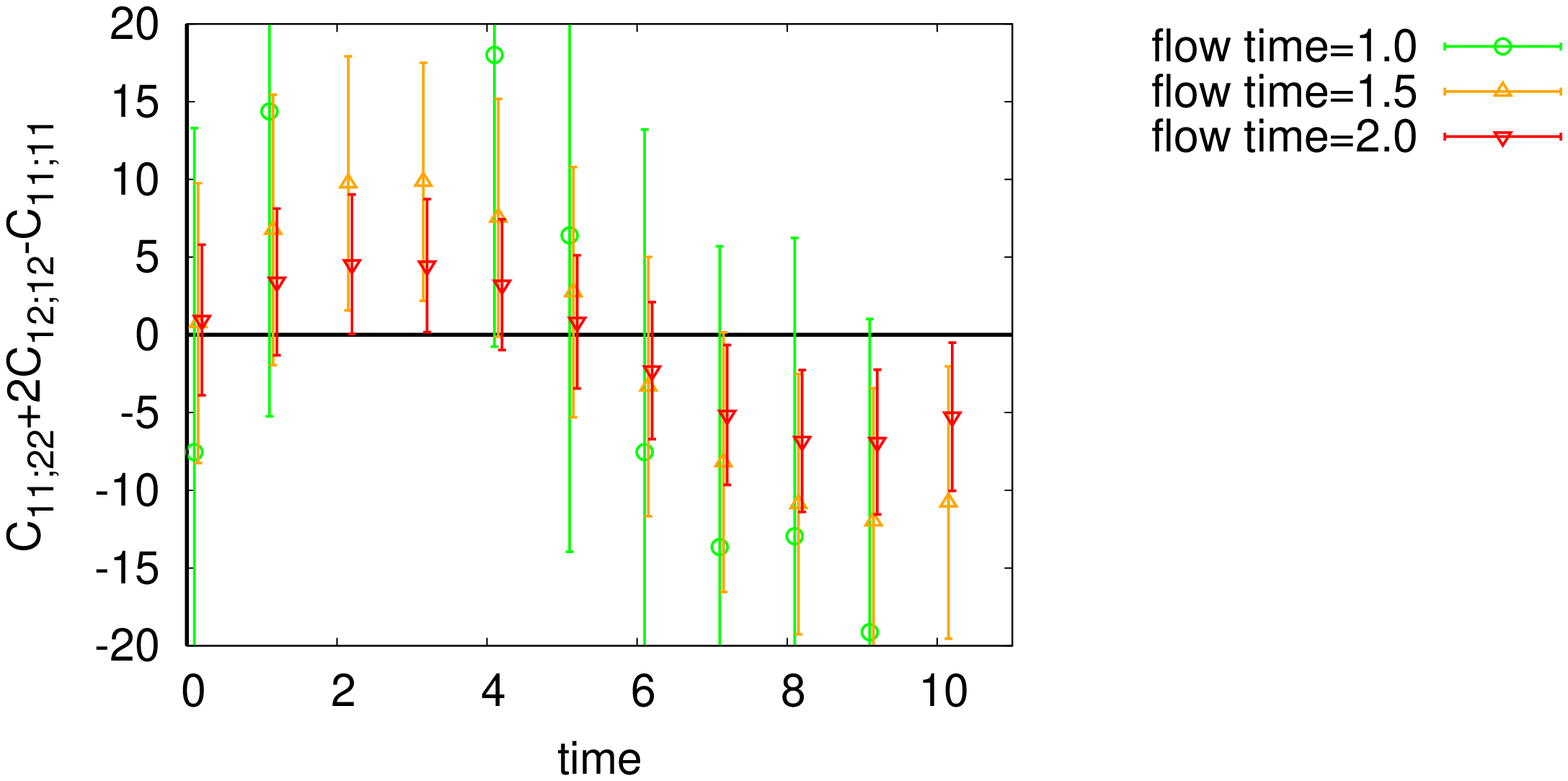}
  \caption{A linear combination of correlation functions 
  $C_{11;22}(\tau)+2C_{12;12}(\tau)-C_{11;11}(\tau)$ as function of the Euclidean time, which should
  vanish if the rotational symmetry is not broken.
  The flow time $t/a^{2}=1.0$ (green circle), $1.5$ (orange up triangle), $2.0$ (red down triangle).}
  \label{fig:rotational}
\end{figure}

Under the three dimension rotational symmetry the spatial component of the correlation function
should take the form
\begin{eqnarray}
C_{ij;kl}(\tau)=c_1(\tau)\delta_{ij}\delta_{kl}
+c_2(\tau)\left(\delta_{ik}\delta_{jl}+\delta_{il}\delta_{jk}\right)
\end{eqnarray}
and correlation functions satisfy the following relation
\begin{eqnarray}
C_{11;22}(\tau)+2C_{12;12}(\tau)-C_{11;11}(\tau)=0.
\label{eqn:rotational}
\end{eqnarray}
In Fig.~\ref{fig:rotational} we plot the left hand side of \eqn{eqn:rotational} as a function of the 
Euclidean time.
The signal becomes better as we proceed the gradient flow.
The linear combination is consistent with zero for all region within $1\sigma$ - $1.5\sigma$.

%----------------------------------------------------------------------------
\subsection{Linear response relations}

In this subsection we test linear response relations using the entropy density.
The entropy density is expressed in three different ways.
First it is written as $(\epsilon+p)/T$ by using the Maxwell's relation and the integrable condition
for the entropy
\begin{eqnarray}
s=\left(\frac{\partial S}{\partial V}\right)_T
=\left(\frac{\partial p}{\partial T}\right)_V
=\frac{\epsilon+p}{T}
=\frac{\langle T_{00}+T_{ii}\rangle}{T},
\label{e+p}
\end{eqnarray}
which is given by the expectation value of one point function of the energy momentum tensor.

On the other hand starting from a linear response of the pressure against a variation of the temperature
the entropy density is given by a derivative of the energy momentum tensor
\begin{eqnarray}
s=\left(\frac{\partial p}{\partial T}\right)_V
=\frac{\partial\langle T_{ii}\rangle}{\partial T}.
\end{eqnarray}
Substituting the statistical physics relation for the expectation value
\begin{eqnarray}
\langle T_{ii}\rangle=\frac{1}{Z}{\rm Tr}\left(T_{ii}e^{-H/T}\right)
\end{eqnarray}
we have the second expression in terms of the two point function of the energy momentum tensor
\begin{eqnarray}
s=\frac{1}{T^2}\langle\delta H \delta T_{ii}\rangle
=\frac{1}{T^2}\int_{V_3}d^3x\left(
 \left\langle \delta T_{00}(\tau,\vec{x})\delta T_{ii}(0)\right\rangle
\right)
=T^{3}C_{00;ii}(\tau).
\label{C00ii}
\end{eqnarray}
The last representation is given by using a linear response to the infinitesimal Lorentz boost
\begin{eqnarray}
\epsilon+p
=\left.\frac{\partial\langle T_{01}\rangle}{\partial v_{1}}\right|_{\vec{v}=0}
=\frac{1}{T}\int_{V_3}d^3x\left(
 \left\langle \delta T_{0i}(\tau,\vec{x})\delta T_{0i}(0)\right\rangle
\right)
=T^{4}C_{0i;0i}(\tau).
\label{C0i0i}
\end{eqnarray}

\begin{figure}[thb] % no figure before 1st section
  \centering
  \includegraphics[width=4.5cm]{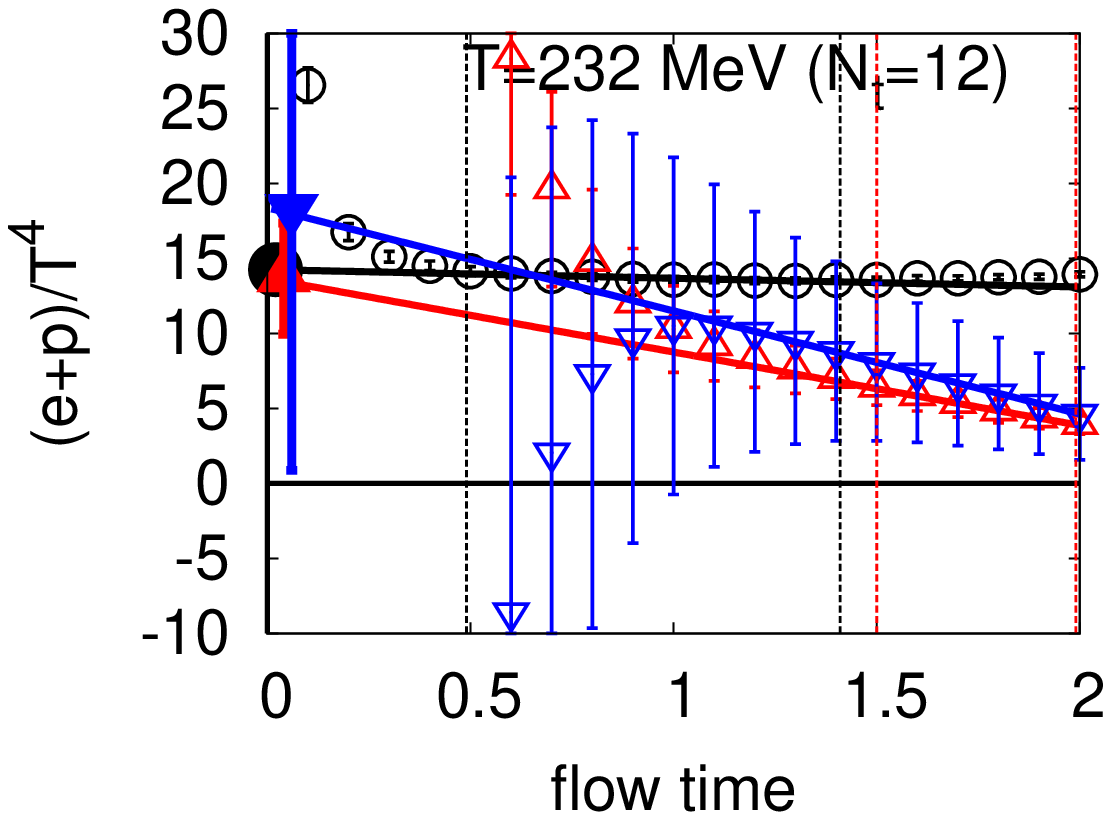}
  \includegraphics[width=4.5cm]{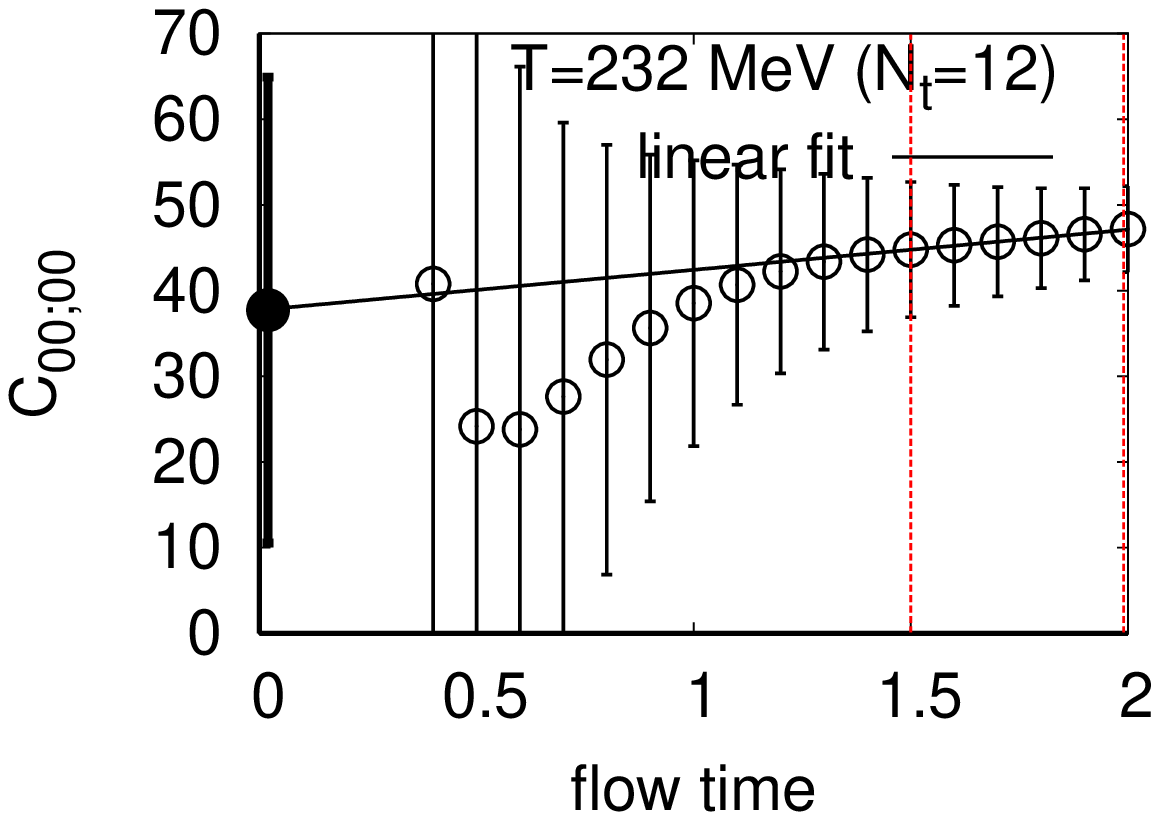}
  \caption{The left panel is three different contributions \eqn{e+p} (black circles),
   \eqn{C00ii} (blue down triangles) and \eqn{C0i0i} (red up triangles)
  to the entropy density plotted as functions of the flow time.
  For $C_{00;22}$ and $C_{02;02}$ we employ three data at $5\le\tau/a\le7$ and fitted by a constant.
  The black and red dotted vertical lines indicate the linear window for the one and two point functions.
  The right panel is $C_{00;00}$ in \eqn{C0000} as a function of the flow time.
  Here again three data at $5\le\tau/a\le7$ are taken and fitted by a constant.
  }
  \label{linear-response}% Give a unique label
\end{figure}

The correlation functions \eqn{C00ii} and \eqn{C0i0i} are plotted as a function of the Euclidean time
in the middle and right panel of Fig.~\ref{fig:conservation} for $i=2$.
According to the discussion given in subsection \ref{sec:conservation} three data at $5\le\tau/a\le7$ 
are employed and fitted by a constant.
The results are plotted in the left panel of Fig.~\ref{linear-response} as a function of the flow time
by blue down triangles  \eqn{C00ii} and red up triangles \eqn{C0i0i}
for the entropy density divided by the temperature $(\epsilon+p)/T^{4}$.
The contribution from the one point function \eqn{e+p} is averaged over space-time and
is plotted by black circles.

After taking the $t\to0$ limit \eqn{eqn-limit} three contributions give the entropy density
$(\epsilon+p)/T^{4}$.
We adopt the same strategy used in Refs.~\cite{Taniguchi:2016ofw,Taniguchi:2016tjc} to take
the limit.
The flowed operator \eqn{eq:(2.8)} evaluated on the lattice behaves as follows as a function
of the flow time
\begin{equation}
 T_{\mu\nu}(t,a) 
  = T_{\mu\nu} +t\,S_{\mu\nu} +A_{\mu\nu}\frac{a^2}{t} + t^2 R_{\mu\nu} ,
\label{eqn:5.4}
\end{equation}
where $A_{\mu\nu}$ appears due to the lattice artifact before taking the continuum limit.
$S_{\mu\nu}$ and $R_{\mu\nu}$ are higher dimensional irrelevant operator.
What we find in our data is the linear window where the nonlinear contributions $a^{2}/t$ and
$t^{2}$ are negligible and data behaves linearly in the flow time.
In the figure the linear window is indicated by the black and red vertical dotted lines for the one point
function \eqn{e+p} and two point correlators \eqn{C00ii}, \eqn{C0i0i}.
The data are fitted linearly in $t$ within the window to take the $t\to0$ limit.

The entropy density given by the linear fit is plotted by the corresponding filled symbols near
the origin.
Three different representations of the entropy density are consistent with each other within the
statistical error.
This indicates a success of the evaluation the energy-momentum tensor correlation functions.

%----------------------------------------------------------------------------
\subsection{Specific heat}

The specific heat is given by a response of the energy against a variation  of the temperature
\begin{eqnarray}
&&
c_{V}=\frac{1}{V_{3}}\frac{d\langle H\rangle}{dT}.
\end{eqnarray}
Inserting the statistical physics relation
\begin{eqnarray}
\frac{1}{V_{3}}\langle H\rangle=\frac{1}{Z}{\rm Tr}\left(T_{00}(x)e^{-H/T}\right)
\end{eqnarray}
it is given in terms of the two point function of the energy momentum tensor
\begin{eqnarray}
c_{V}=\frac{1}{T^{2}}\int_{V_3}d^3x\left(
 \left\langle \delta T_{00}(\tau,\vec{x})\delta T_{00}(0)\right\rangle
\right)
=T^{3}C_{00;00}(\tau).
\label{C0000}
\end{eqnarray}
The correlation function is plotted in the left panel of Fig.~\ref{fig:conservation}.
Here again data at $5\le\tau/a\le7$ are taken and fitted by a constant.
The result are plotted in the right panel of Fig.~\ref{linear-response} as a function of the flow time.
As in the previous subsection we find the linear window indicated by the red vertical lines and
perform the linear fit to take the $t\to0$ limit.
The resultant specific heat is plotted by the filled black circle at the origin, whose value is
$c_{V}=38(27)$.

%----------------------------------------------------------------------------
\section{Conclusion}\label{sec-conclusion}

The two point correlation function of the renormalized energy-momentum tensor is calculated
in $N_{f}=2+1$ QCD on lattice.
Within the statistical error, the correlation function is consistent with a flat behavior in the middle
 region of the Euclidean time, which may be related to the conservation law.
%The correlation function is consistent wth flat within the statistical error in the middle
%apart from the point source, which may be related to the conservation law.
The spatial rotational symmetry is well shown within the statistical error.

The nonperturbative renormalization on lattice is done by applying the gradient flow
to the gluon and quark fields and taking $a\to0$ and then $t\to0$ limits.
Both limits are realized by making use of the linear window and applying the linear fit.
This procedure is tested using the entropy density.
Three difference representations are consistent with each other in the $t\to0$ limit.
Finally we calculate the specific heat by applying the method.

\begin{figure}[thb] % no figure before 1st section
  \centering
  \includegraphics[width=5.4cm]{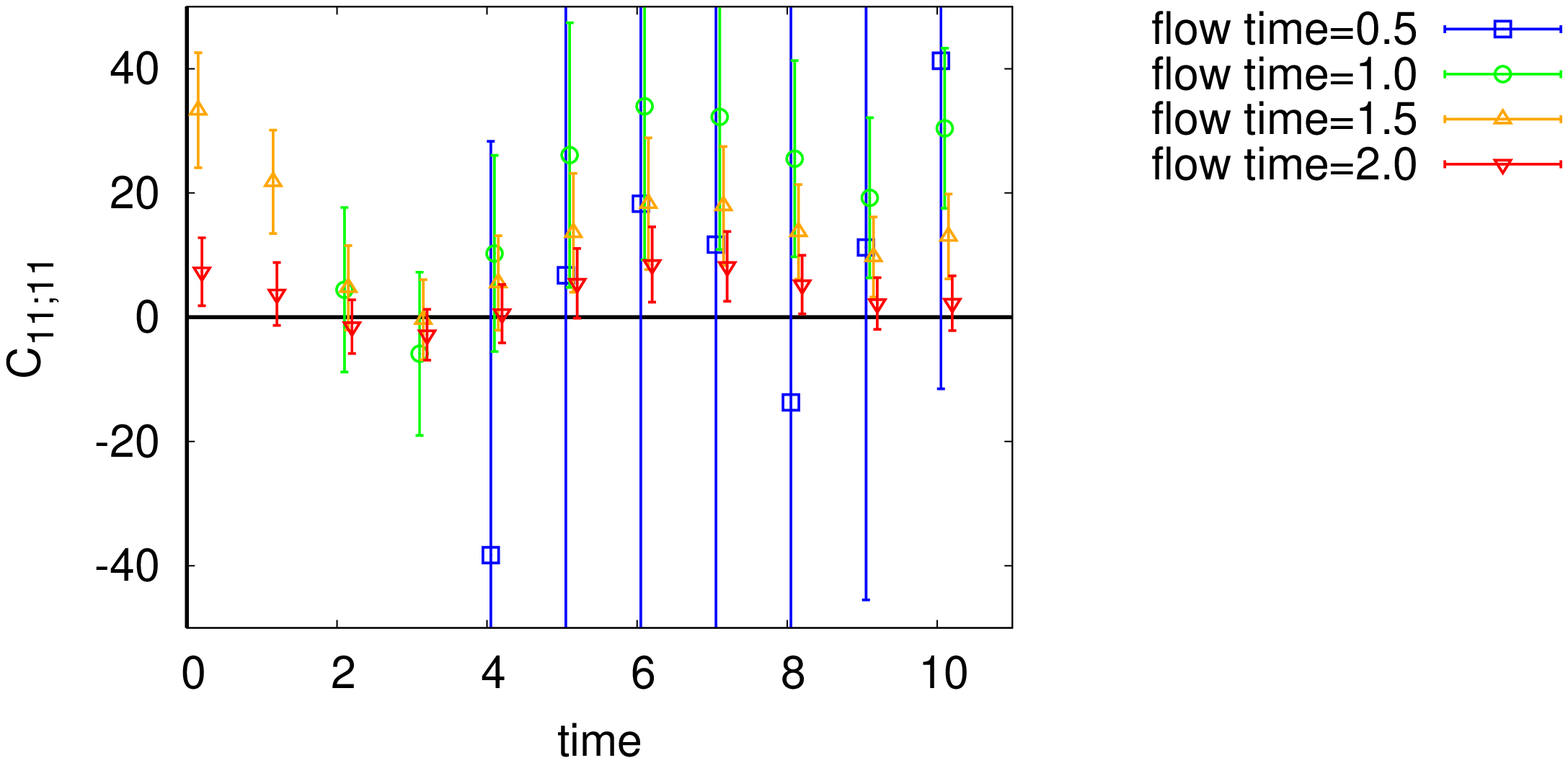}
  \includegraphics[width=5.4cm]{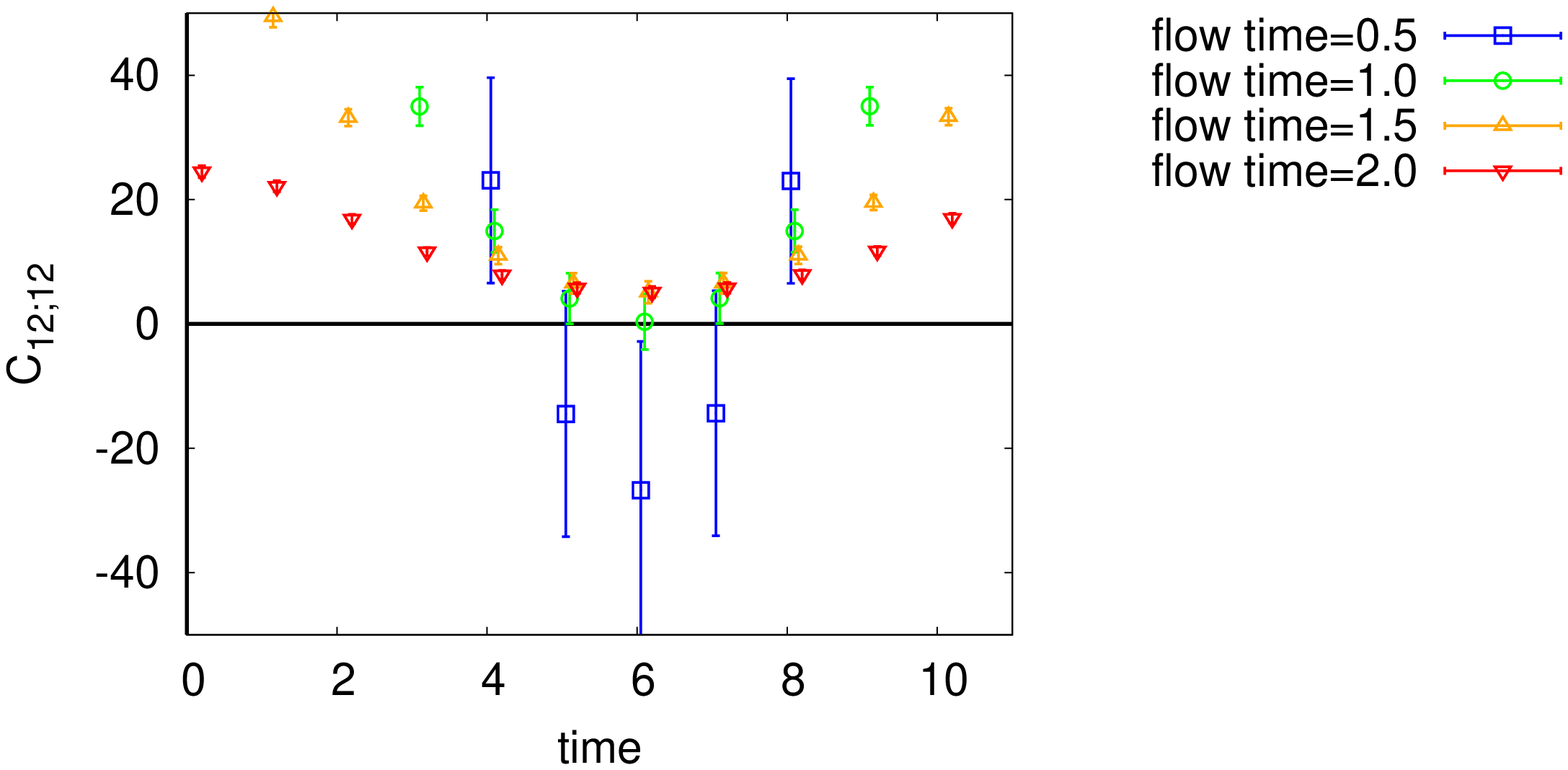}
  \caption{The bulk correlator $C_{11;11}(\tau)$ (left panel) and the shear correlator
  $C_{12;12}(\tau)$ (right panel) as a function of the Euclidean time.
  }
  \label{bulk-shear}
\end{figure}

Future application shall be derivation of the bulk and shear viscosity using the corresponding
two point correlation functions shown in Fig.~\ref{bulk-shear}.
We are also planning to calculate the correlation function on gauge configurations
with physical quark mass, which is used for a derivation of equation of state in Ref.~\cite{Kanaya}.

%----------------------------------------------------------------------------
\vspace{5mm}
%----------------------------------------------------------------------------
%\noindent\textbf{Acknowledgments}

This work was in part supported by JSPS KAKENHI Grant Numbers
JP25800148, JP26287040, JP26400244, JP26400251, JP15K05041, JP16H03982, and JP17K05442.
This research used computational resources of HA-PACS and COMA provided by the Interdisciplinary Computational Science Program of Center for Computational Sciences at University of Tsukuba (No.\ 17a13), 
SR16000 and BG/Q by the Large Scale Simulation Program of High Energy Accelerator
Research Organization (KEK) (Nos.\ 13/14-21, 14/15-23, 15/16-T06, 15/16-T-07, 15/16-25, 16/17-05),
and Oakforest-PACS at JCAHPC through the HPCI System Research project (Project ID:hp170208).
This work was in part based on Lattice QCD common code Bridge++ \cite{bridge}.

%----------------------------------------------------------------------------
%\clearpage
%\bibliography{lattice2017}

%%%%%%%%%%%%%%%%%%%%%%%%%%%%%%%%%%%%%%%%%%%%%%%%%%%%%%%%%%%%%%%%%%%%%%%%%%%%%
\end{document}